# Mid-infrared germanium-on-silicon waveguide sensor for therapeutic drug monitoring of phenytoin

David J. Rowe*[+], Siyu Chen[+], Mihai-Adrian Panainte[^], Callum J. Stirling[+], Pin Dong[o], Monika Bakalarz[+], Hanuushah Vizabaskaran[+], Glenn Churchill[+], Weilin Jin[+], Georgia Mourkioti[+], Martin Ebert[+], Graham T. Reed[+], Saul N. Faust[^], James S. Wilkinson[+], Milos Nedeljkovic[+], Daniel R. Owens[^] and Goran Z. Mashanovich[+].

[+] Optoelectronics Research Centre, University of Southampton, Southampton SO17 1BJ, U.K.

[^] Faculty of Medicine and Institute for Life Sciences, University of Southampton, Southampton SO16 6YD, and NIHR Southampton Clinical Research Facility and Biomedical Research Centre, University Hospital Southampton NHS Foundation Trust, Southampton SO16 6YD, U.K.

[o] School of Physics Engineering and Technology, University of York, Heslington, York YO10 5DD, U.K.

ABSTRACT We report the design, fabrication and characterization of evanescent mid-infrared germanium-on-silicon waveguide sensors for therapeutic drug monitoring (TDM). TDM requires rapid and accurate quantification of serum drug levels but existing clinical assays rely on laboratory-based instrumentation that limits point-of-care deployment. In this work, tunable diode laser absorption spectroscopy was used to analyze dried samples of the anti-seizure medication phenytoin in the spectral region of $\lambda = 5.6 - 6.0$ µm. A limit of detection of 2.20 mg/L was achieved for extracted samples, where phenytoin was first added to human serum and subsequently isolated using liquid-liquid extraction. This limit is significantly below the therapeutic window of 10 – 20 mg/L for phenytoin, enabling detection of sub-therapeutic concentrations. At the same time, the sensor maintains a consistent dose-dependent response up to 40 mg/L, demonstrating its capability to quantify concentrations across the therapeutic window and above the upper therapeutic limit. This validates the use of silicon photonics for biomedical infrared spectroscopy for patients undergoing drug therapy, whether the serum-drug concentration is either too high or too low. These results highlight the potential of mid-IR integrated photonics to form the basis of compact, scalable platforms for point-of-care TDM.

INTRODUCTION Therapeutic drug monitoring (TDM) is the process of measuring the concentration of a medicine from the serum of a patient receiving the drug to ensure safe and effective dosing. It enables a clinician to alter the dose and timing of the drug depending on whether the level is too high or too low. This is particularly important for drugs with a narrow therapeutic window that cause toxicity such as the anti-seizure medication phenytoin, used for the management of epilepsy.

Conventional analytical methods for performing TDM generally use laboratory techniques such as chromatography and mass spectrometry. Such methods require centralized infrastructure and specialist personnel to process samples and operate equipment. These methods have high batch efficiency, where many samples can be processed simultaneously, but systematically introduce delays of several hours for routine tests[1]. Immunoassay kits enable faster turnaround time by testing at the scale of individual patients but suffer from issues with specificity and cross-reactivity[2] while remaining inherent laboratory-based methods because of their microplate format. In contrast, a point-of-care method operable by the non-expert user would enable clinicians to obtain the relevant information directly in hospital, primary care or in the homes of patients. This need for locally delivered testing aligns with current UK government policy driving a shift in NHS services from hospital-based care toward community-centered provision, aimed at improving access and reducing reliance on acute facilities[3].

Optical methods are suitable for point-of-care implementation because the light-sample interaction is more easily automated and miniaturized in comparison with chemical methods. Two such methods are surface-enhanced Raman spectroscopy (SERS) and mid-infrared (mid-IR) absorption spectroscopy.

SERS exploits the highly intense electromagnetic fields of localized surface plasmon resonances for a strong interaction with a target molecule for detection, resulting in low limits of detection in TDM applications[4-6]. However, the reproducibility required for TDM is significantly impacted by the need for the analyte to occupy a nanoscale plasmonic resonance amidst the complexity of a biological matrix[5]. Furthermore, this small overlap makes multiplexing difficult due to competitive adsorption of different drugs onto the sensor surface[6].

Mid-IR absorption spectroscopy using evanescent waveguide sensors has been investigated for applications including protein sensing[7, 8], which could find application in cancer screening[9], and illicit drug monitoring[10]. Cocaine was added to saliva and detected at 500 mg/L concentration using a single-wavelength measurement ($\lambda = 5.8$ μm) with a germanium-on-silicon (GoS) waveguide. This demonstrates that mid-IR photonics could be used for overdose detection in emergency medicine[11], although the relevant limits of detection are much higher than for TDM.

We have chosen to implement mid-IR spectroscopy for TDM using integrated photonics because the platform enables the light-sample interaction to be tuned for appropriate sensitivity while using an interaction volume much larger than that of SERS, thereby avoiding the issues of poor reproducibility and difficult multiplexing. The platform is mass producible due to the fabrication methods of the microelectronics industry, enabling single use transducers to be produced instead of requiring microwell plate-based pipelines associated with immunoassays. Such miniaturized sensors also avoid the need for centralized laboratory infrastructure required for the large and expensive equipment required for LC-MS.

One potential issue with mid-IR spectroscopy of biological fluids is interferents from the biological matrix, in particular water and protein, both of which have broad and intense absorptions in the mid-IR region[12]. In this work, we use a previously-reported extraction method for extracting hydrophobic drugs[13], which removes the water and protein content from the sample and isolates the analyte in a solvent with fewer interfering spectral features. Although the extraction is performed in a laboratory, work is underway to automate this workflow using a microfluidic platform.

TDM is helpful for doctors to maintain the concentration of drugs within a safe and effective treatment range that can be individualized to each patient. Phenytoin is a good example as it has a narrow range between the minimum effective concentration and toxic levels. Individual variability, because of age, genetics, organ function or other medications, can lead to significant shifts in serum concentrations. If the level is too low, seizures might not be controlled; if too high, there is a significant risk of toxicity causing the rapid development of neurological symptoms such as slurred speech and confusion, cardiac arrhythmias and even death, especially in the elderly or critically ill[14]. TDM allows for early identification of ineffective or harmful drug levels before symptoms occur. The therapeutic window of phenytoin is 10 – 20 mg/L[15].

METHODS Rib waveguides were designed for the GoS material platform using simulations in Lumerical (Ansys, USA). This material platform was used because of its broad transparency range. A parameter sweep of the rib width was used to design the waveguide for single mode operation across the wavelength range of the tunable laser source, $\lambda = 5.5 – 11$ μm. Because a range of rib widths satisfied this condition, the maximum width of 3.0 μm was used to give the greatest modal confinement, which would reduce sidewall interaction and the corresponding propagation loss. The waveguide height was 3.2 μm and the etch depth was 1.8 μm. Tapered edge couplers and waveguide bends were also simulated in Lumerical. The taper design was optimized for minimal coupling loss, widening from the waveguide width to 16 μm at the facets with a taper length of 3 mm. For waveguide bends, excess propagation loss is inversely proportional to bend radius so a minimum bend radius of 300 μm was chosen as a compromise between the compactness of the circuit and a manageable excess bend loss. The simulated excess bend loss was 0.1 dB per 90° bend for TM polarization and 0.54 dB per 90° bend for TE polarization. TM polarization was used for its increased evanescent field interaction with an analyte dried on the waveguide surface.

A schematic diagram of the chip layout is shown in Figure 1, where the blue shaded areas represent etched regions. A nested array of fifteen rib waveguides was arranged so all waveguides had the same total length and number of 90° bends for the same propagation and coupling losses. A fluid confinement region was designed so that a trench encircled part of the waveguide array, which aimed to limit the spread of a droplet deposited at its center. Each waveguide had a different optical interaction length contained within the confinement region.

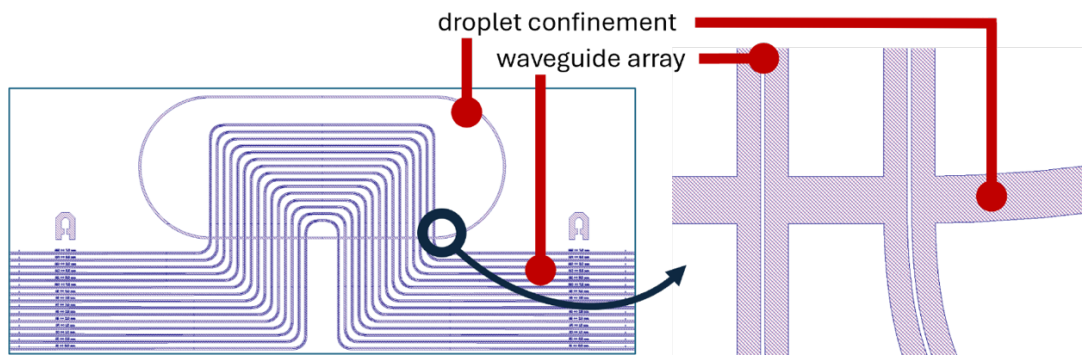

Figure 1: schematic of the photonic integrated circuit showing the waveguide array being partially encircled by the fluid confinement trench. The blue areas represent regions that were etched into the Ge layer; white areas represent unetched Ge.

Devices were fabricated using 200 mm GoS wafers as part of a CORNERSTONE multi-project wafer fabrication run at the University of Southampton. Features were patterned using deep UV lithography and etched using inductively coupled plasma-reactive ion etching. The waveguide ribs and fluid confinement trench were etched together in a single etch step. The etched wafers were diced into chips using ductile dicing, which yielded facets of sufficient optical quality, including low roughness, low topside chipping, high flatness and high verticality, to enable end-fire coupling into the tapered couplers[16].

A schematic of the measurement setup used to perform absorption spectroscopy is shown in Figure 2. Absorption measurements were performed with a MIRcat tunable quantum cascade laser (Daylight Solutions, USA), operating in continuous-wave mode across the wavelength range 5.6 – 6.0 μm. This range was selected to be wide enough to cover the phenytoin absorption peak of interest[13] while including narrow zero-absorption regions on either side to enable reliable baseline correction. At the same time, the scan range was sufficiently narrow to allow rapid wavelength sweeping of the laser source. The average output power of the laser was around 450 mW (Class 3R), which posed a potential risk of damaging both the detector and the FLIR Boson mid-IR camera (FLIR, USA). To mitigate this, wedged neutral density (ND) filters were placed after the laser to attenuate the optical power. The wedged geometry further helped suppress unwanted back-reflections within the optical system.

Since the rib waveguides were designed to operate under TM polarization for high surface interaction, a linear polarizer was placed after the ND filters to ensure high polarization purity of the incident beam. The beam was then focused onto the waveguide input taper using a reflective objective lens with 40× magnification and a numerical aperture of 0.5, providing a tightly focused spot and improved coupling efficiency into the waveguide. At the output side of the waveguide, a second reflective objective lens was used to collect the transmitted light and direct it to the mid-IR camera. The image from the mid-IR camera allows verification that the QCL laser excites the intended guided mode of the rib waveguide rather than undesired slab modes. Once correct mode excitation and alignment were confirmed, a foldable mirror was used to redirect the transmitted beam to a HgCdTe photodetector (VIGO Photonics, Poland), which exhibits a typical detectivity $D^*$ of ~$4.0 \times 10^9$ cm·Hz$^{1/2}$/W at its peak response, and the optical alignment was further optimized to maximize the detected power for spectral measurements.

All optical components were arranged in close proximity to avoid unnecessary beam propagation in air and so minimize the interaction length with environmental water vapour. This approach suppressed background absorption features to ensure that the measured spectrum was dominated by the waveguide-analyte interaction rather than environmental contributions. The total free space pathlength was approximately 120 cm.

For the spectral measurements, after the optimized alignment was found, the laser was swept linearly across the 5.6 – 6.0 μm wavelength range at a constant scan rate of 0.1 μm s$^{-1}$, resulting in a deterministic mapping between time and wavelength. Under this condition, the recorded detector signal represents a time-domain encoding of the optical absorption spectrum. To suppress high-frequency noise arising from electronic fluctuations and residual interference fringes while preserving the intrinsic absorption line shape, each individual wavelength scan was processed with a low-pass filter (LPF) procedure. The filter cut-off frequency was selected based on the narrowest absorption features of interest in this spectral region (on the order of 5 nm), corresponding to a characteristic temporal frequency of approximately 20 Hz at the applied scan rate. Accordingly, a cut-off frequency of 40 Hz was chosen to ensure full preservation of the absorption profile while reducing the high frequency noise. Filtering was implemented using a fifth-order Butterworth filter with zero-phase forward-backward processing to avoid phase distortion or artificial broadening of absorption features. To further improve the signal-to-noise ratio, the wavelength scan was repeated 100 times under identical experimental conditions. The filtered spectra from all scans were then averaged to reduce uncorrelated noise while maintaining reproducibility of the measured absorption features. This result was treated as a single measurement. The total insertion loss of the TDM chip was measured to be 16.2 dB at a wavelength of 5.85 μm using a typical chip.

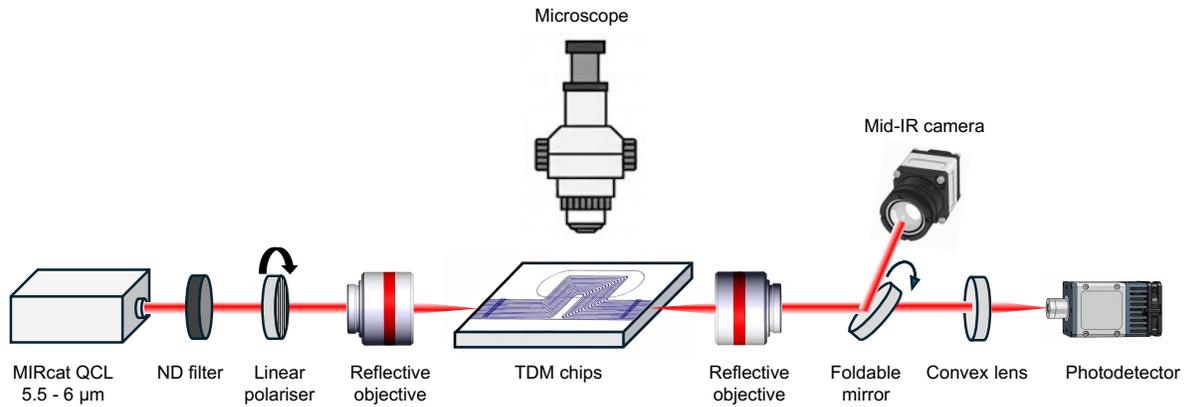

Figure 2: schematic of the measurement setup used to perform absorption spectroscopy. The TDM chips were mounted on a three-axis stage for alignment.

Phenytoin (≥ 98% purity; Cayman Chemical, US) was prepared in human serum (male AB plasma, USA origin, sterile filtered; Merck, UK) at concentrations of 5, 10, 20, and 40 mg/L. The solutions were subsequently extracted into ethyl acetate following a previously published method[13]. A blank serum control underwent the same extraction procedure but without the addition of phenytoin.

The multistep extraction was used to remove major endogenous interferents from the serum solutions and can be summarized as follows. Lipoproteins were first precipitated by adding a magnesium chloride–dextran sulfate sodium solution, enabling removal of lipid-bound components. The supernatant was then treated with saturated ammonium sulfate to salt-out albumin, which retains the protein-bound phenytoin; the supernatant containing small water-soluble molecules was discarded. The albumin precipitate was resuspended in water and extracted twice with ethyl acetate to isolate phenytoin while removing remaining proteins. The organic extracts were combined and the solvent was evaporated, and the dried residue was reconstituted in ethyl acetate for subsequent analysis.

For each measurement, a reference spectrum was recorded by first measuring the power transmission spectrum of each waveguide. In all cases, this reference corresponds to the clean waveguide transmission prior to any droplet deposition. The same referencing procedure was applied for both analyte-containing samples and for the blank serum control.

0.5 μL of sample was then pipetted onto the waveguide array and allowed to dry under ambient conditions. In practice, drying took several seconds. The waveguides were measured a second time in the same way to provide sample spectra. Absorbance was calculated as $10 \log_{10}$ of the ratio of the sample and reference spectra. All samples were measured in triplicate; that is, aliquots of each sample were pipetted onto three waveguide arrays and characterized individually. A new waveguide array was used to characterize each aliquot.

The noise of the system was estimated using the standard deviation of repeated measurements of the blank serum control. To achieve this, the entire measurement procedure, where 100 scans were filtered and averaged for a single spectrum, was repeated twenty times for each of three aliquots of the blank serum control for every waveguide.

RESULTS To determine the limit of detection (LoD) and observe dependence on waveguide interaction length, four waveguides were characterized for extracted phenytoin samples. Waveguide 6 had interaction length $l = 3.44$ mm, waveguide 8 $l = 4.44$ mm, waveguide 10 $l = 5.44$ mm and waveguide 12 $l = 6.44$ mm. Each waveguide had a total length of 11.79 mm.

Figure 3 shows the absorbance spectra for multiple concentrations of extracted phenytoin, measured using waveguides with different interaction lengths. The solid line represents the mean spectrum and the shaded areas represent ± one standard deviation, taken from the repeated measurements of each sample. Interference bands associated with water vapor from the free-space path of the experimental setup are shown as gray shaded regions.

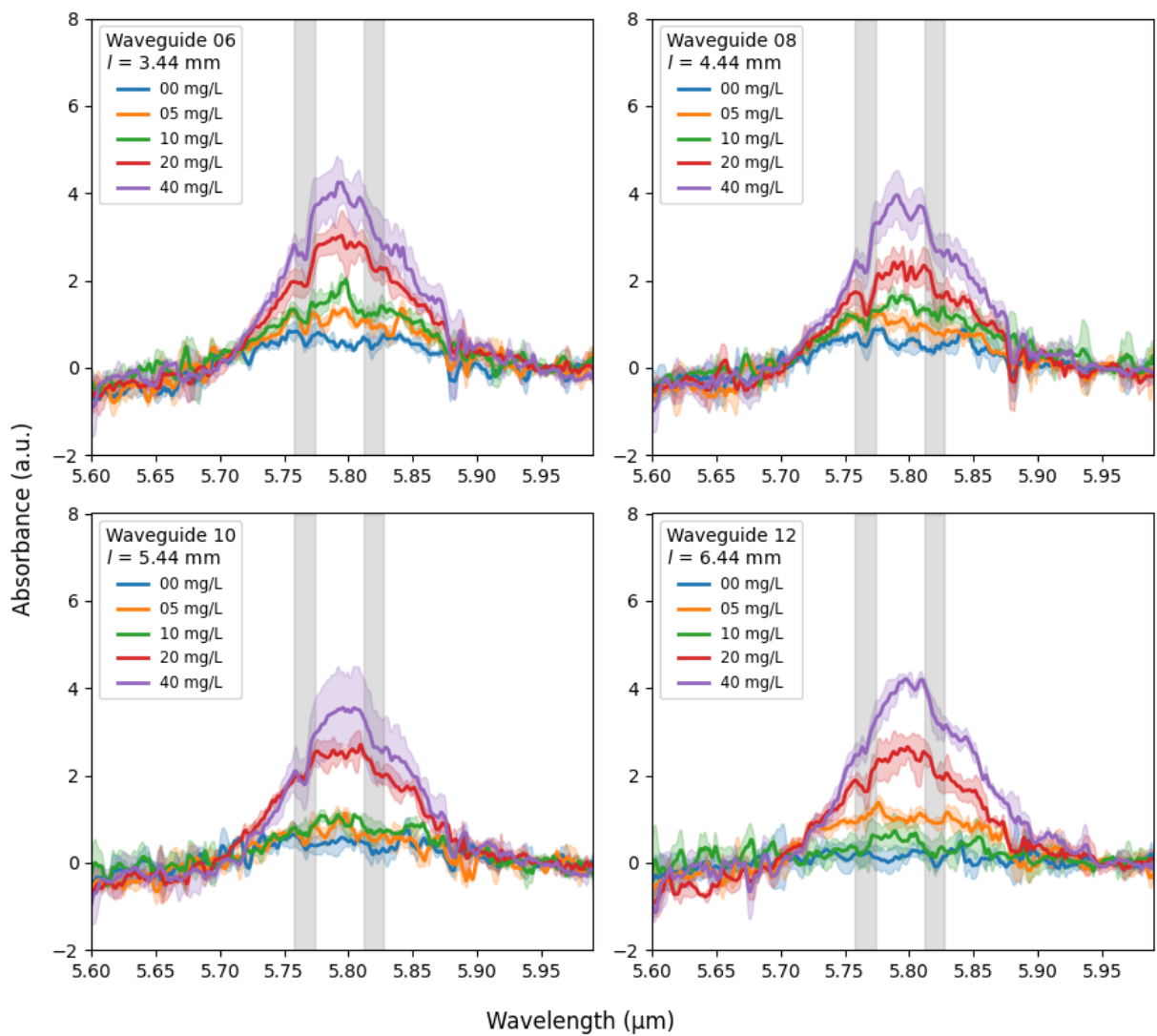

Figure 3: absorbance spectra for multiple concentrations of extracted phenytoin. Each subplot represents a different waveguide corresponding to a different interaction length *l*. Error bars represent ± one standard deviation from measurements of multiple aliquots. The zero-concentration trace represents blank measurements where the extraction protocol was applied to serum that had not had any phenytoin added. The gray shaded regions show interference from atmospheric water vapor and are excluded from the subsequent AUC calculation.

Figure 4 shows the result of integrating the areas under the curves (AUC) in Figure 3 within the wavelength region 5.71 – 5.87 μm, covering the main phenytoin absorption peak, plotted with respect to concentration and again plotted for waveguides with different interaction lengths. Water vapor interference bands, shown as gray shaded regions in Figure 3, were excluded from the integration. For each subplot in Figure 4, the height of the shaded area represents the 3σ noise limit, calculated from the AUC from the repeated blank measurements, where three aliquots of the blank sample were measured 20 times each. The LoD was calculated as the concentration corresponding to this 3σ noise limit using the fit line.

Because the blank underwent the full extraction protocol, a small amount of residual organic material remained on the waveguide surface after drying, giving rise to a weak but reproducible broadband absorbance background. This background is not subtracted; therefore, it appears as a non-zero y-intercept in the calibration curve shown in Figure 4. It is for this reason that the 3σ noise limit is taken from the y-intercept in order to extract the LoD.

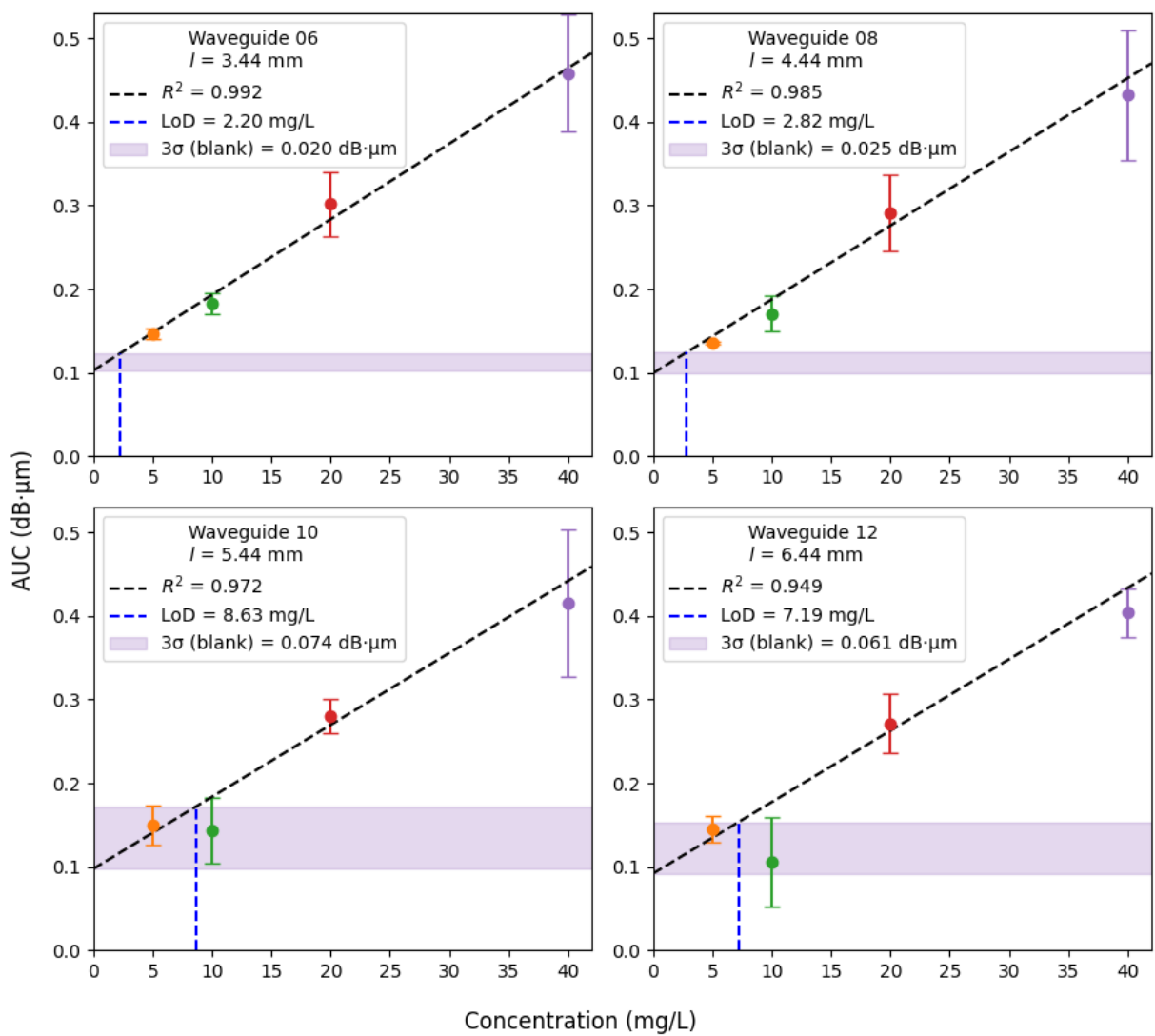

Figure 4: AUC with respect to concentration for extracted phenytoin. Each subplot represents a different waveguide corresponding to a different interaction length. Error bars represent ± one standard deviation from measurements of multiple aliquots. The height of the shaded area represents the 3σ noise limit of the blank measurement. The limit of detection (LoD) was calculated as the concentration corresponding to an AUC equal to the 3σ noise limit above the y-intercept of the calibration curve.

Figure 4 shows that the LoD reached is as low as 2.20 mg/L for the shortest waveguide and below 9 mg/L for all waveguides, all below the therapeutic range of 10-20 mg/L. No saturation of the concentration response is seen throughout this range, and above 20 mg/L the concentration response continues to show linearity up to 40 mg/L.

DISCUSSION Figure 4 shows that the optimum performance for detecting extracted phenytoin in terms of limit of detection, sample error and measurement noise is obtained using the shortest two interactions lengths (waveguides 6 and 8). The best LoD is 2.20 mg/L, which is significantly lower than the therapeutic window of 10 – 20 mg/L for phenytoin and suggests that this method would be appropriately sensitive for TDM in the clinically most likely cases where a patient is underdosed. The concentration response shows linear behavior up to at least 40 mg/L, within the accuracy of the measurements, with the zero-concentration intercept being offset by 0.1 dB·µm due to non-specific absorption from trace organic material, salts and residual solvent that remain after the extraction process, as seen in the unspiked sample (0 mg/L curves in Fig 3). While the low LoD is relevant for identifying sub-therapeutic concentrations, the response up to 40 mg/L shows that normal and supra-therapeutic levels can also be quantified, enabling control of dosing to bring the concentration within the therapeutic range.

The improved performance compared with previously reported mid-IR waveguide sensors for biofluid analysis likely arises from two factors. First, the strong evanescent field overlap in GoS rib waveguides provides high interaction strength with the dried analyte film. Second, the extraction protocol removes the dominant mid-IR interferents in serum, including water and proteins, substantially reducing background absorption. Together, these effects lead to a higher signal-to-noise ratio and enable clinically relevant limits of detection.

The measured 3σ noise of the blank sample is dominated by system-level effects and is greater than would be expected from the noise-equivalent power of the detector. Instead, the dominant noise sources arise from laser-related fluctuations, including relative intensity noise and wavelength jitter during rapid QCL sweeping, together with variations in the effective waveguide interaction length caused by sample drying and wicking. These effects are further aggravated by residual atmospheric water vapor absorption over the free-space optical path. Accordingly, the LoD is determined by these combined system-level contributions.

The sample error increases systematically with concentration for the concentration responses with the best coefficients of correlation (R). This scaling of the error with concentration is most likely due to variations in sample preparation or in the nature of the sample drying process.

The gradient of the fit line is virtually identical for all waveguides, while it was expected that an increased interaction length would yield a larger gradient for a given concentration. This implies that the waveguide length over which the sample dries is no different despite different lengths being enclosed within the confinement trench. While preliminary experiments using these TDM chips show that the liquid confinement approach adopted is suitable for aqueous samples with high contact angles and strong droplet pinning, it appears much less effective at controlling the spread of liquids with low contact angles such as ethyl acetate. It is deduced that here the sample "wicks" along the waveguides outside the confinement trench. This could be avoided by depositing a mid-IR transparent cladding layer on the chip and etching a window in the cladding to expose the desired length of waveguide. In this way, the sample boundary would be precisely set by design rather than by droplet shape. Completely flooding such a cavity would naturally enhance sample uniformity because it would introduce pinning at the cavity edge, suppressing radial capillary flow and making evaporative flux more uniform. The appropriate material and processes are under development.

The LoD is determined by the 3σ measurement error on the blank and the gradient of the AUC-concentration response. It can be observed that the 3σ measurement error on the blank, visually represented by the height of the shaded area in Fig 4, generally increases with waveguide design length. This error on the blank is believed to be dominated either by (i) the variability in the effective path length due to the "wicking" of the sample along the waveguide, which would be reduced by introducing a cladding as described above, or by (ii) fluctuations in the laser power, which could be addressed by inclusion of an on-chip splitter to divide the input into separate sensing and referencing waveguides. These could either be measured simultaneously using two balanced detectors or measured sequentially using a single detector by recombining into a single waveguide and switching between the two[17]. Further reduction in LoD could be achieved by including a longer waveguide length in the sample, which would increase the gradient of the AUC-concentration curve and so decrease the LoD for a given error performance.

These results demonstrate that mid-IR silicon photonics has appropriate sensitivity for TDM of phenytoin in serum and point towards several avenues for improvement.

CONCLUSION This work reports the design and validation of evanescent GoS waveguide sensors for TDM of phenytoin using spiked serum samples. The limit of detection of 2.20 mg/L is lower than the therapeutic window of 10 – 20 mg/L, enabling identification of sub-therapeutic concentrations, while the linear concentration-dependent response up to 40 mg/L demonstrates the ability to quantify supra-therapeutic levels. Together, these results show that mid-IR GoS waveguides provide both the sensitivity and dynamic range required for clinically relevant therapeutic drug monitoring.

The study also highlights practical considerations for optimizing device performance, including control of sample spreading and interaction length. These aspects offer clear pathways for further refinement, for example through the use of straight waveguides and integrated cladding layers to define the sample region precisely. Future work will focus on extending the wavelength range and validating the approach using patient-derived samples containing metabolites and additional drugs, building towards a robust mid-IR photonic platform for point-of-care TDM.

AUTHOR INFORMATION

**Corresponding Author**

* Correspondence should be addressed to Dr David J. Rowe at d.rowe@soton.ac.uk.

**Author Contributions**

The manuscript was written through contributions of all authors. All authors have given approval to the final version of the manuscript.

ACKNOWLEDGMENT

The authors acknowledge funding from the Engineering and Physical Sciences Research Council (grant no. EP/V047663/1).

ABBREVIATIONS

QCL, quantum cascade laser; ATR-FTIR, attenuated total reflection-Fourier transform infrared.